\documentclass{elsart}
\usepackage{epsfig}
\usepackage{amsmath}
\usepackage{amssymb}
\usepackage{graphicx}
\usepackage{subfigure}

\newcommand \be{\begin{equation}}
\newcommand \ba{\begin{eqnarray}}
\newcommand \ee{\end{equation}}
\newcommand \ea{\end{eqnarray}}

\begin{document}
\runauthor{W.-X. Zhou \& D. Sornette}
\begin{frontmatter}
\title{Antibubble and Prediction of China's stock market and Real-Estate}
\author[iggp]{\small{Wei-Xing Zhou}},
\author[iggp,ess,nice]{\small{Didier Sornette}\thanksref{EM}}
\address[iggp]{Institute of Geophysics and Planetary Physics, University of California, Los Angeles, CA 90095}
\address[ess]{Department of Earth and Space Sciences, University of California, Los Angeles, CA 90095}
\address[nice]{Laboratoire de Physique de la Mati\`ere Condens\'ee, CNRS UMR 6622 and Universit\'e de Nice-Sophia Antipolis, 06108 Nice Cedex 2, France}
\thanks[EM]{Corresponding author. Department of Earth and Space
Sciences and Institute of Geophysics and Planetary Physics,
University of California, Los Angeles, CA 90095-1567, USA. Tel:
+1-310-825-2863; Fax: +1-310-206-3051. {\it E-mail address:}\/
sornette@moho.ess.ucla.edu (D. Sornette)\\
http://www.ess.ucla.edu/faculty/sornette/}

\begin{abstract}
We document a well-developed log-periodic power-law antibubble in
China's stock market, which started in August 2001. We argue that
the current stock market antibubble is sustained by a contemporary
active unsustainable real-estate bubble in China. The
characteristic parameters of the antibubble have exhibited
remarkable stability over one year (Oct. 2002-Oct. 2003). Many
tests, including predictability over different horizons and time
periods, confirm the high significance of the antibubble
detection. We predict that the Chinese stock market will stop its
negative trend around the end of 2003 and start going up,
appreciating by at least 25\% in the following 6 months.
Notwithstanding the immature nature of the Chinese equity market
and the strong influence of government policy, we have found maybe
even stronger imprints of herding than in other mature markets.
This is maybe due indeed to the immaturity of the Chinese market
which seems to attract short-term investors more interested in
fast gains than in long-term investments, thus promoting
speculative herding.

\end{abstract}

\end{frontmatter}

\section{Introduction} \label{s1:intro}

After the Third Plenary Session of the 11th Central Committee of
the Communist Party of China in December 1978, China adopted new reform
policies and committed to open to the outside. These reforms have favored
China an unprecedented growth in Gross Domestic Product (GDP). If
we normalize the GDP index (deflated by the Consumer
Price Index (CPI) for inflation) to 100 in 1978, then the GDP
index grew to 781.2 in 2001 \cite{CSYB02}: China's economy has
grown about 9\% per year on average. In the past two decades,
China has experienced a transition from a centrally planned
economy to a market economy. During this period, one of the most
important development has been the setup of the Chinese stock market, which
has played an increasingly important role in the transition of the economy.

Before the foundation of People's Republic of China in 1949, the
Shanghai stock exchange was the third largest worldwide, after New
York and London and its evolution over the period from 1919 to
1949 had enormous influence on other world-class financial markets
\cite{Su03}. After 1949, China implemented policies of a socialist
planned economy and the government controlled entirely all
investment channels. This proved to be efficient in the early
stage of the economy reconstruction, especially for the heavy
industry. However, planned economic policies have unavoidably led
to inefficient allocation of resources. In 1981, the central
government began to issue treasury bonds to raise capital to cover
its financial deficit, which reopened the China's securities
markets. After that, local governments and enterprises were
permitted to issue bonds. In 1984, 11 state-owned enterprises
became share-holding corporations and started to provide public
offering of stocks. The establishment of secondary markets for
securities occurred in 1986 when over-the-counter markets were set up
to trade corporation bonds and shares. The first market for
government-approved securities was founded in Shanghai on
November 26, 1990 and started operating on December 19 of the same year
under the name of the Shanghai Stock Exchange (SHSE). Shortly after, the
Shenzhen Stock Exchange (SZSE) was established on December 1, 1990
and started its operations on July 3, 1991.

Thanks to the establishment of the SHSE and SZSE, the Chinese
stock market\footnote{By ``Chinese stock market'', we refer to
that in the mainland of China.} has grown rapidly. In 1991, the
total market capitalization was 284.4 billion yuan (13.2\% of GDP)
with the float market capitalization being 85.1 billion yuan
(3.94\% of GDP). In 2002, the total market capitalization
was 4,033.7 billion yuan in total (39.4\% of GDP) with 1,167.4
billion of float market capitalization (11.4\% of
GDP)\footnote{Since 1995, the exchange rate of {\it{Renminbi}}
(RMB) yuan has been pegged to the US dollar with small fluctuations
allowed around 8.27 yuan per US dollar.}. The historical high happened in 2000
when the total market capitalization reached 4,968 billion yuan
(55.5\% of GDP) with 1,535.4 billion yuan of float market
capitalization (17.2\% of GDP). The size of the Chinese stock
market has increased remarkably.


There are different types of China-related stock shares, including
several tradable shares (state owned shares, legal person shares
and employee shares) and untradable shares (A-shares, B-shares,
H-shares, red chips and other foreign shares including N-shares
and S-shares).
A mainland Chinese company qualifying for
equity issuance has to keep about 67\% of its capital in the form
of state-owned shares and legal person shares and can only issue
up to 33\% of its capital in the form of A- and/or B- and/or
H-shares and/or employee shares during initial public offering
(IPO) \cite{Su03}. Only A-shares and B-shares are traded in the
SHSE and SZSE. Despite of the separation of A- and B-shares, their
daily (log) returns are correlated as a result of the information
transmission mechanism at work \cite{CK98JFR}. There are many
problems in the emerging Chinese stock market which include (i)
small scale, (ii) instability, (iii) untradability of more than
2/3 of the shares, (iv) absence of shorting, (v) significant
impact of government policies, (vi) over-speculation, (vii)
over-valuation of the markets, (viii) widely taken short-term
positions, (ix) insider trading, and (x) distempered regulation
system \cite{C03}. These specific features make the market exhibit
strong idiosyncracies and puzzles in addition to the more common
behavior of mature stock markets \cite{FR98SSRN}. The foreign
shares has been found to behave differently from domestic shares
in several respects \cite{X00CER}. The commonly reported day of
the week effects \cite{B86,JL00} are found to be absent in both
Shanghai and Shenzhen A and B markets \cite{P01CAFR}. The
acceptance or rejection of the random-walk hypothesis in the
Chinese stock markets is controversial, and the conclusion depends
on the approach used to perform the tests
\cite{LSR97AEL,DZ00FR,Su03}. The markets also have extremely high
volatility \cite{Su03} and very high P/E (price-over-earning)
ratios \cite{SH02FB,SZ02FB}. The high P/E ratios can be partially
explained by the relative scarcity of shares resulting from
the holding of more than 2/3 of their total
number by the government.

While, as we briefly summarized, there are many problems that can
distort the analysis of the Chinese stock markets, it is an
interesting question to consider the possibility for behaviors
similar to those which have been documented for other developed
markets \cite{S03,S03PR}, in particular in view of the documented
speculative nature of Chinese investors \cite{C03}. We refer in
particular to the US and worldwide ``antibubble'' regime found to
have developed after the collapse of the so-called information and
internet bubble in 2000 \cite{SZ02QF,ZS02Global}. We find several
puzzles in the behavior of the Chinese stock markets in the last
few years. The Chinese stock markets turned into a bearish
antibubble regime (significantly decreasing prices)
\footnote{Since we have documented in the past two types of
antibubbles, one bullish and the other bearish
\cite{ZS02Global,J03QF,SZ03QF,GP03Polish}, we refer to the bearish
antibubble type when we use the term ``antibubble'' in this paper
as it is the only relevant in the present study.} in asynchrony
with most of the major world markets \cite{ZS02Global} and
significantly later at a time when China's economy was still doing
well. The SHSE Composite Index culminated at the all-time high of
2242.4 on 2001/06/13 and since plummeted 32.2\% to 1520.7 on
2001/10/22. It reached its minimum at 1319.9 on 2003/01/03, with
many strong preceding local ups and downs. Notwithstanding the
asynchrony with the rest of the world, we will show below that the
evolution of the Chinese stock markets since mid-2001 can be quite
well described by the same concept of an antibubble.

The paper is organized as follows. In
Sec.~\ref{s1:datasets}, we describe the data sets. In
Sec.~\ref{s1:PLLP}, we identify separately two signatures of an
antibubble characterizing the
trajectory of the SHSE Composite Index:
(i) a power-law relaxation and (ii) log-periodic undulations.
A combined log-periodic
power-law (LPPL) analysis in Sec.~\ref{s1:LPPLfits} further
confirms the existence of an antibubble in the Chinese stock market. We investigate the
statistical properties of the fit residuals in
We test in Sec.~\ref{s1:predict} the
predictive potential of our characterization of an antibubble regime.
Section \ref{s1:UnivIdio}
compares the 2001 China antibubble with other documented
antibubbles and describe how the antibubble was gestated and how it
developed. Finally, Sec.~\ref{s1:concl} concludes.

\section{Data sets}
\label{s1:datasets}

There are many different indexes published by SHSE, SZSE and other
investment companies. These indexes can be decomposed into four categories:
composite indexes, sample share indexes, classified indexes and other
indexes. We have retrieved fourteen indexes produced by three
organizations, that is, the SHSE, the SZSE and TX Investment Co.
Ltd. whose indexes are also well-known and widely adopted by
investors\footnote{See http://www.txsec.com}. These indexes are
the SHSE Composite Index (SHSE\_CI), SHSE 180 Index (SHSE\_180), SHSE
A Share Index(SHSE\_A), SHSE B Share Index (SHSE\_B), SZSE
Composite Index (SZSE\_CI), SZSE A Share Subindex (SZSE\_SA), SZSE
B Share Subindex (SZSE\_SB), TX 280 Index (TX\_280), TX Total
Share Index (TXT), TX SHSE Total A Share Index (TXT\_SH\_A), TX
SZSE Total A Share Index (TXT\_SZ\_A), TX Float Share Index (TXF),
TX SHSE Float A Share Index (TXF\_SH\_A), and TX SZSE Float A
Share Index (TXF\_SZ\_A).

\begin{figure}
\begin{center}
\includegraphics[width=7cm]{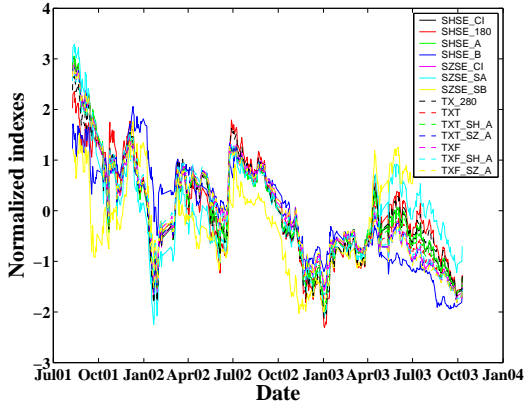}
\end{center}
\caption{(Color online) Comparison of 14 normalized indexes:
SHSE Composite Index (SHSE\_CI), SHSE 180 Index (SHSE\_180), SHSE
A Share Index(SHSE\_A), SHSE B Share Index (SHSE\_B), SZSE
Composite Index (SZSE\_CI), SZSE A Share Subindex (SZSE\_SA), SZSE
B Share Subindex (SZSE\_SB), TX 280 Index (TX\_280), TX Total
Share Index (TXT), TX SHSE Total A Share Index (TXT\_SH\_A), TX
SZSE Total A Share Index (TXT\_SZ\_A), TX Float Share Index (TXF),
TX SHSE Float A Share Index (TXF\_SH\_A), and TX SZSE Float A
Share Index (TXF\_SZ\_A).}
\label{Fig:AllInd}
\end{figure}

The SHSE and SZSE indexes are from 2000/08/09 to 2003/10/28 and
the TX indexes are from 2000/08/09 to 2003/10/22. The normalized
times series of these indexes to zero mean and unit variance are
drawn in Fig.~\ref{Fig:AllInd}. One can see that, apart from the
two B-share-related indexes SHSE\_B and SZSE\_SB which exhibit
significant deviations, the other twelve indexes show quite
similar trajectories with relatively minor differences. In this
paper, we shall present our results based on the SHSE Composite
Index but all our conclusions remain valid for other eleven
indexes with similar behavior.

\section{Power law and log-periodicity}
\label{s1:PLLP}

The assumption that there is a critical point at the inception of
an antibubble can be tested by investigating market behavior
close to criticality. Two possible signatures of a singular
or critical behavior might appear during
the development of an antibubble: a power law relaxation and
log-periodic wobbles. In this section, we
analyze these two features separately. A combined log-periodic
power-law analysis is presented later in Sec.~\ref{s1:LPPLfits}.

\subsection{Power law relaxation}
\label{s2:PL}

The power law expression for an antibubble reads
\begin{equation}
 I(t) = A + B (t-t_c)^m~,
 \label{Eq:PL}
\end{equation}
where $t_c$ is an estimate of the inception of the antibubble,
$t>t_c$, and the order parameter $I(t)$ can be the price
$p(t)$ or its logarithm $\ln[p(t)]$. We shall discuss which of these
two choices is the more pertinent for $I(t)$ in the next
section. If $m<0$, $I(t)$ is singular when $t \to t_c^+$ and $B$
should be found positive to ensure
that $I(t)$ decreases. If $0<m<1$, $I(t)$ is finite but
its first derivative $I'(t)$ is singular at $t_c$ and $B$ should
be found negative to ensure that $I(t)$ decreases.

\begin{figure}[h]
\begin{center}
\includegraphics[width=7cm]{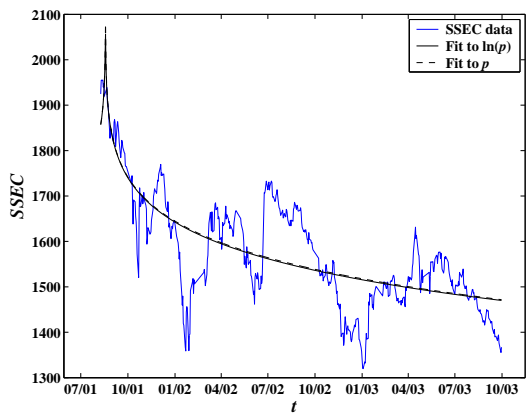}
\end{center}
\caption{Daily trajectory of the SHSE Composite Index
from 2001/08/09 to 2003/10/10 (wiggly line) and fits using a power
law formula (\ref{Eq:PL}). The continuous line is the fit to the
logarithms of the SHSE Composite Index ({\it{i.e.}}, $I(t) =
\ln[p(t)]$), while the dashed lines show fits to the SHSE
Composite Index ({\it{i.e.}}, $I(t) = p(t)$).} \label{Fig:FitPL}
\end{figure}

We fit the SHSE Composite Index from 2001/08/09 to 2003/10/10
using the power law formula (\ref{Eq:PL}). The fit with
$I(t)=\ln[p(t)]$ is illustrated in Fig.~\ref{Fig:FitPL} as a solid
line, whose parameters are $t_c={\rm{2001/08/18}}$, $m=0.15$,
$A=7.76$, and $B=-0.1691$ with a r.m.s. of the fit residuals
$\chi=0.0519$. The dashed line in Fig.~\ref{Fig:FitPL} gives the
fit for $I(t)=p(t)$, whose parameters are $t_c={\rm{2001/08/18}}$,
$m=0.10$, $A=2526$, and $B=-531.7$ with a r.m.s. of the fit
residuals $\chi=80.2$. The two fits estimate a same inception date
of the antibubble and gives close power law exponents. The two
fits are too close to be distinguishable in Fig.~\ref{Fig:FitPL}.

\subsection{Log-periodicity \label{s2:LP}}

The detection of log-periodic oscillations, in any, is
conveniently performed by removing the global trend of the index.
One way is to subtract the power law fit (\ref{Eq:PL}) from the
index and by analyzing the wobbles of the obtained residue $s(t)$
by an adequate spectral analysis. We shall also use a
non-parametric approach called the $(H,q)$ analysis. These two
methods provide a residue $s(t)$ and we stress that their results
are independent of a priori assumptions of the presence of
log-periodicity, a welcome property in view of some criticisms
concerning a previous implementation \cite{JSL99JR} of such
analysis \cite{F01QF}.

Since log-periodicity corresponds to regular oscillations in the
variable $\ln (t-t_c)$, we use a Lomb periodogram analysis which
is well-adapted to the uneven sampling of the variable $\ln
(t-t_c)$ \cite{Press}. A Lomb analysis also allows us to assess
the statistical significance level of the extracted
log-periodicity \cite{Press,ZS02IJMPC,B03QF}.

\subsubsection{Parametric detrending approach \label{s3:paradet}}

\begin{figure}[h]
\begin{center}
\includegraphics[width=7cm]{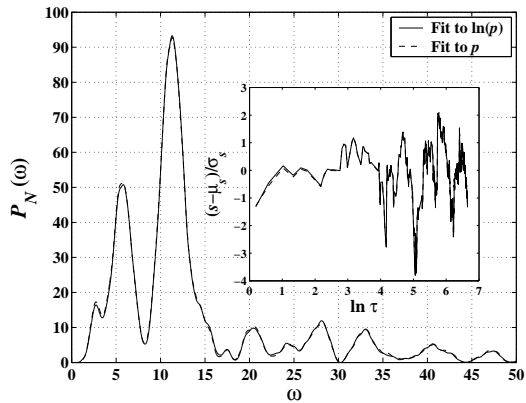}
\end{center}
\caption{Lomb periodogram analysis of $s(t)$ where $s(t)$ is
considered to be a function of $\ln\tau$ where $\tau = t-t_c$.
In other words, $\omega$ is the log-angular frequency conjugate
to $\ln\tau$; thus a strong significant peak of the periodogram
qualifies the existence of log-periodicity. The two choices
 $I(t)=\ln[p(t)]$ and $I(t)=p(t)$ have been considered and give
the same undistinguishable results. The inset shows the centered
normalized residue
$[s(t)-\mu_s]/\sigma_s$ as a function of $\ln \tau$. Note that
$P_N(\omega)<10$ for $\omega>50$ so that the part of the spectrum
shown here exhibits the most relevant part.} \label{Fig:Lomb}
\end{figure}

Following \cite{JSL99JR,JLS00IJTAF}, the first method consists of
removing the power law trend measured in Sec.~\ref{s2:PL}.  We
construct the residue $s(t)$ in the following way
\begin{equation}
s(t) = [I(t)-A]/(t-t_c)^{m}~, \label{Eq:Res}
\end{equation}
where $A$, $m$ and $t_c$ are obtained for the fit of the pure power law
formula (\ref{Eq:PL}) to the data, as shown in Sec.~\ref{s2:PL}. This
residue $s(t)$
has a nonzero mean $\mu_s$ and the variance $\sigma_s^2$ different from $1$.
The inset of Fig.~\ref{Fig:Lomb} plots $[s(t)-\mu_s]/\sigma_s$ as a
function of $\ln \tau$, where we denote
\be
\tau = t-t_c~.
\ee
The Lomb periodogram of this centered normalized
residue considered as a function of the variable  $\ln \tau$ is shown in
Fig.~\ref{Fig:Lomb}. Again, the two choices $I(t)=\ln[p(t)]$ and
$I(t)=p(t)$ are undistinguishable. Since $P_N(\omega)$ is a
normalized Lomb power, $s(t)$ and $[s(t)-\mu_s]/\sigma_s$ have
identical Lomb periodogram.

The highest peak for $I(t)=\ln[p(t)]$ ($I(t)=p(t)$) is at
$\omega=11.29~(11.31)$ with height $P_N =
93.3~(92.7)$. This gives a false alarm probability\footnote{False
alarm probability is the chance that we falsely detect
log-periodicity in a signal without true log-periodicity.}
${\rm{Pr}}\ll 10^{-5}~(10^{-5})$ for white noise \cite{Press}. If
the residues have long range-correlations characterized
by a Hurst index $H>1/2$, we have ${\rm{Pr}}
< 10^{-5} $ for $H=0.6$, ${\rm{Pr}}=0.001$ for $H=0.7$,
${\rm{Pr}}=0.02$ for $H=0.8$, and ${\rm{Pr}}=0.10$ for $H=0.9$.
The second highest peak is at $\omega=5.82~(5.74)$ with $P_N =
50.8~(51.2)$. We interpret these two highest peaks as a
fundamental log-periodic component at $\omega_f = 5.82~(5.74)$ and
it harmonic component at $2\omega_f$.

\subsubsection{$(H,q)$-analysis \label{s3:HqA}}

We have also performed a generalized $q$-analysis, called $(H,q)$-analysis,
on $I(t)$, which is a non-parametric method for characterizing
self-similar functions. The $(H,q)$-analysis \cite{ZS02PRE,ZS04IJMPC}
is a generalization
of the $q$-analysis \cite{E97PLA,EE97PRL}, which is a natural tool
for the description of discretely scale invariant fractals. The
$(H,q)$-derivative is defined as
\begin{equation}
D_q^H I(\tau) \stackrel{\triangle}{=} \frac
{I(\tau)-I(q\tau)}{[(1-q)\tau]^H}~. \label{Eq:HqD}
\end{equation}
The special case $H=1$ recovers the standard $q$-derivative, which
itself reduces to the standard derivative in the limit $q \to 1^-$.
There is no loss of generality by constraining $q$ in the open
interval $(0,1)$ \cite{ZS02PRE}. The advantage of the
$(H,q)$-analysis is that there is no need for detrending, as it
is automatically accounted for by the finite difference and
the normalization by the denominator.

We apply the $(H,q)$-analysis to $I(x) = \ln p(t)$
to check for the existence of
log-periodicity. The results obtained for
$I(t)=p(t)$ are very similar. The independent variable
is taken to be $\ln \tau$, with $t_c$ equal to
the value $t_c={\rm{2001/08/18}}$ estimated above from the
simple power law fit (\ref{Eq:PL}) \cite{ZS02PRE}.
The same method has been applied to test for
log-periodicity in stock market bubbles and antibubbles
\cite{SZ02QF,ZS04IJMPC}, in the USA foreign capital inflow bubble
ending in early 2001 \cite{SZ03inflow}, and in the ongoing UK real
estate bubble \cite{ZS03PhysA}.

\begin{figure}[h]
\begin{center}
\includegraphics[width=7cm]{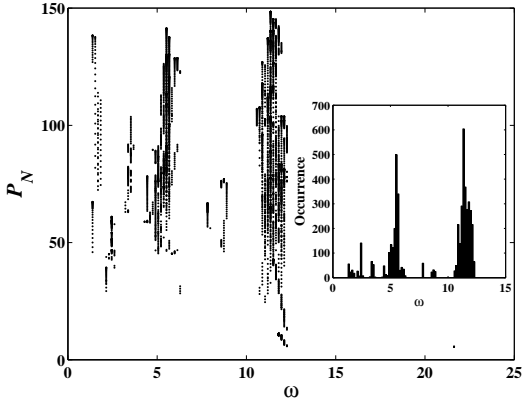}
\end{center}
\caption{Bivariate distribution of pairs $(\omega,P_N)$ defined as the
highest Lomb peaks of the $(H,q)$-derivative of $I(t)=\ln[p(t)]$.
Each point in the figure corresponds the highest Lomb peak and its
associated angular log-frequency in the Lomb periodogram of the $(H,q)$-derivative
of $I(t)=\ln[p(t)]$ for a given pair $(H,q)$. The inset shows the
marginal distribution of $\omega$ in the interval $0 < \omega , 13$.}
\label{Fig:HqA}
\end{figure}

For each pair of $(H,q)$ values, we
calculate the $(H,q)$-derivative, on which we perform a Lomb
analysis. We find that most of the Lomb periodograms have a
shape similar to that shown in Fig.~\ref{Fig:Lomb}. The highest Lomb peak of
the resultant periodogram has height $P_N$ and
abscissa $\omega$ with height, both $P_N$ and $\omega$ being
functions of $H$ and $q$. We scan a grid of $100
\times 50$ rectangular in the $(H,q)$ plane,
with  $H = -0.99:0.02:0.99$ and $q =
0.01:0.02:0.99$. Figure~\ref{Fig:HqA} shows the bivariate distribution of
pairs $(\omega,P_N)$. The inset gives the marginal distribution of
$\omega$'s in the interval $o < omega <15$. The two most prominent
clusters correspond respectively to $\omega_f=5.42 \pm 0.34$,
identified as the fundamental angular log-frequency and to
$2\omega_f = 11.50\pm 0.38$ interpreted as its harmonics.
We also find a total of five pairs with
$\omega>13$. All these five pairs have the
same $\omega=21.42$, which is very close to
$4\omega_f$. This suggests that that the first, second and fourth
harmonics are expressed in the signal. However, the fourth
harmonic has a weak spectral amplitude $P_N = 5.5 \sim 5.6$.
Other smaller clusters are most probably due to noise
decorating the power law \cite{HJLSS00IGR} or due to a residual
global trend in the $(H,q)$-derivative. In summary, this $(H,q)$-analysis
provides even stronger evidence for the existence of
log-periodicity than the parametric detrending approach
of the previous section \ref{s3:paradet}.

\section{A combined log-periodic power law analysis of the
2001 Chinese antibubble} \label{s1:LPPLfits}

We have shown that, since August 2001, the SHSE Composite Index qualifies
as a critical antibubble according to two signatures of a
power law relaxation (no strong discriminant) decorated by
log-periodic oscillations (very discriminant). In
this section, we perform a combined parametric analysis of these
two signatures, using a log-periodic power law (LPPL) formulation.

\subsection{Fit with first-order LPPL formula}
\label{s2:LPPLfit1}

This LPPL formulation is based on the following principles.
Investors in the stock market form a small-world network
\cite{M67PT,W03} and interact with each other ``locally'' by
imitation. Local interactions propagate spontaneously into global
cooperation leading to herding behaviors, which result in bubbles
and antibubbles. These ingredients together with an assumption
that prices reflect the imitative properties of the system can be
captured by rational imitation models of bubbles and antibubbles
\cite{JSL99JR,JLS00IJTAF,JS00IJTAF}. The main consequence of the
model is that the dynamics may evolve towards or away from
a critical point or critical time $t_c$ corresponding to the most
probable end of the bubble or beginning of the antibubble.
The competition between nonlinear trend followers and nonlinear value
investors together with inertia between investor decisions and
their market impact may lead to additional nonlinear oscillations
approximating a log-periodicity \cite{Idesor1,Idesor2}. Log-periodicity may also result
or be amplified by the existence of a naturally existing hierarchy
of social group sizes \cite{JLS00IJTAF,JS00IJTAF,Zhou_Hill}.

Mathematically, the LPPL structure can be expressed as:
\begin{equation}
I\left( t\right) = A + B \tau ^{\alpha} +C \tau^\alpha \cos\left[
\omega \ln \left( \tau \right) - \phi \right], \label{Eq:fit1}
\end{equation}
where $\phi$ contains two ingredients: the information on the
mechanism of interactions between investors and a rescaling of time
\cite{J03QF,SZ03QF}. The distance to the critical
time is $\tau=t_c-t$ for bubbles and $\tau=t-t_c$ for antibubbles.

Theoretically, the order parameter $I(t)$ can be the price $p(t)$ or
the log of price $\ln[p(t)]$, depending on the following criterion.
Let us assume that the observed price is
the sum $p(t) = F(t) + M(t)$ of a fundamental price $F(t)$ and of a bubble or an
antibubble $M(t)$. We have $I(t) = p(t)$
when $F\ll M$ and $I(t)=\ln[p(t)]$ when $F\sim M$
\cite{JS99_Risk,JS00IJTAF}. The difficulty is that only $p(t)$ is observed
and one does not know how to decompose it into a fundamental contribution
and a bubble or antibubble component.
A statistical test using the distributions of
the $\omega$ and $m$ can be used to determine the relevant order
parameter \cite{JS00IJTAF} but is not always discriminant. For the lack of a better
argument, we try both possibilities and we fit the Shanghai Stock Exchange
Composite index from 2001/08/09 to 2003/10/10 using
(\ref{Eq:fit1}) with either $I(t)=p(t)$ or $I(t)=\ln[p(t)]$.
It turns out that the two cases give results that are
undistinguishable, as in the above analyses. We follow the
numerical algorithm described in Ref.~\cite{SZ02QF}.

\begin{figure}[h]
\begin{center}
\includegraphics[width=7cm]{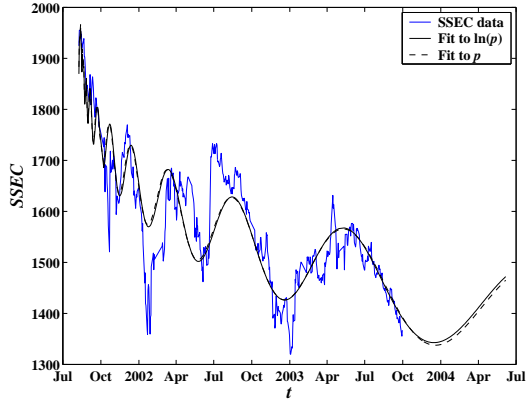}
\end{center}
\caption{Daily trajectory of the SHSE Composite Index
from 2001/08/09 to 2003/10/10 (wiggly line) and fit using the LPPL
formula (\ref{Eq:fit1}). The continuous line is the fit to the
logarithms of SHSE Composite Index ({\it{i.e.}}, $I(t) =
\ln[p(t)]$), while the dashed
line shows the fit to the SHSE Composite Index ({\it{i.e.}}, $I(t) =
p(t)$). Both fits are almost undistinguishable.} \label{Fig:Fit1}
\end{figure}

Figure~\ref{Fig:Fit1} shows the fit of the
SHSE Composite Index (wiggly line) by the LPPL formula (\ref{Eq:fit1})
(continuous line) using $I(t) = \ln[p(t)]$. The fit parameters are
$t_c={\rm{2001/08/12}}$, $m=0.28$, $\omega=11.45$, $\phi=1.77$,
$A=7.62$, $B=-0.0517$, $C=-0.0097$, and the r.m.s. of the fit
residuals is $\chi=0.0385$. We have used calendar days as unit of time.
At least five LPPL oscillations are clearly visible
in the figure.

Using $I(t) = p(t)$, we find the following parameters
$t_c={\rm{2001/08/12}}$, $m=0.23$,
$\omega=11.35$, $\phi=1.14$, $A=2084$, $B=-138.3$, $C=-20.5$, and
the r.m.s. of the fit residuals is $\chi=60.6$, corresponding to the
dashed line shown in Fig.~\ref{Fig:Fit1}. The two fits
produce almost the same trajectory and basically the same
extrapolation. Both fits give very
close values for $t_c$, $\omega$ and $m$. In the rest of this
paper, we shall thus mostly use $I(t)=\ln[p(t)]$, as there is
no need to duplicate almost identical results with $I(t) = p(t)$.

\subsection{Fit with second-order Weierstrass-type function}
\label{s2:LPPLfit2Wei}

While the LPPL formula (\ref{Eq:fit1}) is able to synchronize with
the major oscillations of the Shanghai Stock Exchange Composite
index, there are significant discrepancies in amplitudes. This
should not be too surprising as expression (\ref{Eq:fit1}) is just
the first term in a systematic Fourier series of harmonics. The
Lomb periodograms discussed above have shown the importance of
second and higher-order harmonics. Actually, the existence of a
systematic series of log-periodic terms is expected on general
grounds from the mathematical formulation of discrete scale
invariance in terms of a discrete renormalization group map $R$
such that ${\tau}' = R({\tau})$ and ${\mathcal{F}}({\tau}) =
{\mathcal{G}}({\tau}) + \frac{1}{\mu}{\mathcal{F}}[R({\tau})]$
\cite{SJB96JPI}, where ${\mathcal{F}}({\tau})=\ln[p(t)]-
\ln[p(t_c)]$ such that $\mathcal{F}=0$ at the critical point at
$t_c$, ${\tau}=|t-t_c|$ is the distance to the critical point
occurring at $t_c$, and $\mu>1$ describes the scaling of the index
evolution upon a rescaling of time. These equations can be solved
in the form of an infinite series which can be re-summed by
applying the Mellin transform to exhibit the singular part
\cite{GS02PRE}. This formalism suggests therefore to add at least
a second harmonic to (\ref{Eq:fit1}). More terms can be taken into
account to describe Weierstrass-type functions as in
Refs.\cite{GS02PRE,ZS03RG,ZS03PhysA}. The corresponding expression
generalizing formula (\ref{Eq:fit1}) reads
\begin{equation}
I(t) = A + B {\tau}^{m} + {\Re{\left(\sum_{n=1}^N C_n
{\rm{e}}^{i\psi_n}{\tau}^{-s_n}\right)}}~. \label{Eq:fit2}
\end{equation}
This formula (\ref{Eq:fit2}) with $N=2$ proved to
outperform our former specification in terms of a first-order LPPL
function (\ref{Eq:fit1}) for the modelling of the worldwide
2000-2002 anti-bubbles that started in
mid-2000 \cite{SZ02QF,ZS02Global,GP03Polish} and of the 1975-2001
bubble in the American foreign assets capital inflow \cite{SZ03inflow}.

Fitting the SHSE Composite Index from 2001/08/09 to 2003/10/10 for
$I(t)=\ln[p(t)]$ with expression (\ref{Eq:fit2})
with $N=2$ gives the following parameter values:
$t_c={\rm{2001/08/06}}$, $m=0.12$,
$\omega=5.83$, $\phi_1=2.33$, $\phi_2=6.20$, $A=7.89$,
$B=-0.2692$, $C= 0.0180$, $D= 0.0253$, and the r.m.s. of the fit
residuals is $\chi=0.0319$. For $I(t)=p(t)$, we get
$t_c={\rm{2001/08/03}}$, $m=0.06$, $\omega=6.09$, $\phi_1=3.84$,
$\phi_2=0.00$, $A=3300$, $B=-1233.3$, $C=41.8$, $D=-53.2$, and the
r.m.s. of the fit residuals is $\chi=50.0$. It is worth noting
that the fact that $|C|<|D|$ implying a stronger second harmonic
component is consistent with the result in
Fig.~\ref{Fig:Lomb} showing a higher peak at $2\omega_f$.
Note that the inclusion of the second harmonic is able to account well for
the major features of the SHSE Composite Index trajectory,
while higher frequency effects, such as jumps and crashes, are
still poorly described and would require higher order terms.

\begin{figure}[h]
\begin{center}
\includegraphics[width=7cm]{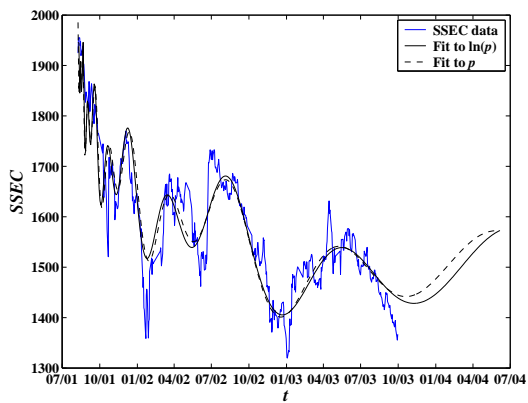}
\end{center}
\caption{Daily trajectory of the SHSE Composite Index
from 2001/08/09 to 2003/10/10 (wiggly line) and fits using a
second-order Weierstrass-type LPPL formula (\ref{Eq:fit2}). The
continuous line is the fit to the logarithms of the SHSE Composite
Index ({\it{i.e.}}, $I(t) = \ln[p(t)]$), while the dashed line shows
the fit to the SHSE Composite
Index ({\it{i.e.}}, $I(t) = p(t)$).} \label{Fig:Fit2}
\end{figure}

As shown in Fig.~\ref{Fig:Fit2}, the two fits are very close with
consistent future scenarios. A comparison of
the first- and second-order fits shows that they are remarkably
consistent in the sense that both fits give close estimates of
the critical time $t_c$, a small power law exponent $m$, and the same fundamental
angular log-frequency $\omega \approx 6$. The existence of a
fundamental angular log-frequency and its high-order harmonics is
a strong signal of the significance of the log-periodicity. We
shall elaborate further on this point later.

\subsection{Fit with second-order Landau expansion function}
\label{s2:LPPLfit2Landau}

Another extension to the log-periodic function (\ref{Eq:fit1}) is
to allow for nonlinear coupling between harmonics, by rewriting
the fundamental existence of a critical exponent in terms of the
following Landau expansion \cite{SJ97PhysA}:
\begin{equation}
 \frac{dI({\tau})}{d\ln {\tau}} = \alpha I({\tau})+\beta |I({\tau})|^2I({\tau})~...~,
 \label{Eq:Landau0}
\end{equation}
where the coefficients may be generally complex. Starting with the
Landau expansion in the vicinity of the critical point $\tau=0$, the first order
term recovers Eq.~(\ref{Eq:fit1}), while the second order gives:
\begin{equation}
I(t) = A + \frac{B{\tau^m}+ C{\tau^m}\cos\left[\omega\ln(\tau) +
\frac{\Delta_{\omega}}{2m}\ln \left(1+\left(\frac{\tau}
{\Delta_t}\right)^{2m}\right) +\phi\right]} {\sqrt{1+
\left(\frac{\tau} {\Delta_t}\right) ^{2m}}}~. \label{Eq:Landau2}
\end{equation}
Higher orders in the Landau expansion add additional degrees
of freedom and allow a more precise description \cite{JS99IJMPC},
however at the cost of more numerous parameters to fit. The value
of a given order should always be determined by the tradeoff between
a better fit and the parameter cost.

Expression (\ref{Eq:Landau2}) has an approximate power
law envelop and describes a transition
from a log-periodicity with angular log-frequency $\omega$ for
$\tau<\Delta_t$ to a log-periodicity with
angular log-frequency $\omega+\Delta_\omega$ for $\Delta_t<\tau$. For
instance, the 1990 Nikkei antibubble experienced the transition from
the first-order Landau description (\ref{Eq:fit1}) to the
second-order Landau formula (\ref{Eq:Landau2})
2.5 years after the inception of the antibubble
\cite{JS99IJMPC,JS00IJMPC}. In contrast, the 2000 S\&P 500 antibubble has
just entered the second-order Landau regime, on the fourth quarter of 2003
after three years \cite{SZ02QF,ZS0310}. Using the LPPL Landau expansion
up to the third order, a prediction was published in
January 1999 on the behavior of the Japanese stock market in the
following two years \cite{JS99IJMPC}, which has been remarkably
successful \cite{JS00IJMPC}.

\begin{figure}[h]
\begin{center}
\includegraphics[width=7cm]{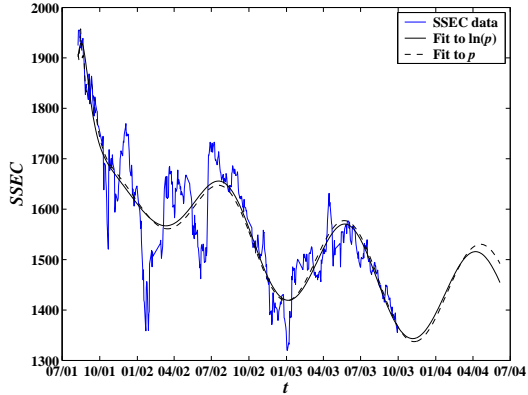}
\end{center}
\caption{Daily trajectory of the SHSE Composite Index
from 2001/08/09 to 2003/10/10 (wiggly line) and fits using a
second-order Landau expansion LPPL formula (\ref{Eq:Landau2}). The
continuous line is the fit to the logarithm of the SHSE Composite
Index ({\it{i.e.}}, $I(t) = \ln[p(t)]$).
The dashed lines show the fit to the SHSE Composite
Index ({\it{i.e.}}, $I(t) = p(t)$).} \label{Fig:FitLandau}
\end{figure}

We fit the SHSE Composite Index to the second-order Landau
formula (\ref{Eq:Landau2}) as shown in
Fig.~\ref{Fig:FitLandau}. The fit parameters for $I(t)=\ln[p(t)]$
are $t_c={\rm{2001/08/12}}$, $\alpha=0.37$, $\omega=6.36$,
$\phi=1.09$, $\Delta{t}=3807$, $\Delta{w}=-93.3$, $A=7.62$,
$B=-0.0323$, $C= 0.0066$, and the r.m.s. of the fit residuals is
$\chi=0.0363$. The fit parameters for $I(t)=p(t)$ are
$t_c={\rm{2001/08/15}}$, $\alpha=0.36$, $\omega=5.89$,
$\phi=0.21$, $\Delta{t}=4757$, $\Delta{w}=-99.9$, $A=1990$,
$B=-54.3$, $C=-10.8$, and the r.m.s. of the fit residuals is
$\chi= 56.7$.

In contrast with the fits with (\ref{Eq:fit1}) or (\ref{Eq:fit2}),
expression (\ref{Eq:Landau2}) smoothes out the oscillations and
other fluctuations at early times of the antibubble but captures
adequately the later oscillations with correct amplitude. This
phenomenon was also observed in the Nikkei case where the
oscillations in the first two months disappeared when using
(\ref{Eq:Landau2}) \cite{JS99IJMPC}, suggesting that they lack
significance. As $\Delta_t$ is more than ten years, this would
lead to conclude that the transition expected for the angular
log-frequency to the second order regime has not yet started.
However, this conclusion is not completely warranted in the
present case in view of the large values found for $\Delta{w}$,
which implies a more complex crossover from the second-order
landau expansion to the first order as $\Delta_t$ increases.

\section{Predictive power of LPPL formulae \label{s1:predict}}

\subsection{When was the antibubble detectable?}
\label{s2:Detactable}

As we have shown that the Chinese stock markets antibubble started
in August 2001, an interesting question arises naturally:
how long after its inception would we have been
able to detect the antibubble? This question amounts to
testing the stability of the antibubble.

\begin{figure}[h]
\begin{center}
\includegraphics[width=7cm]{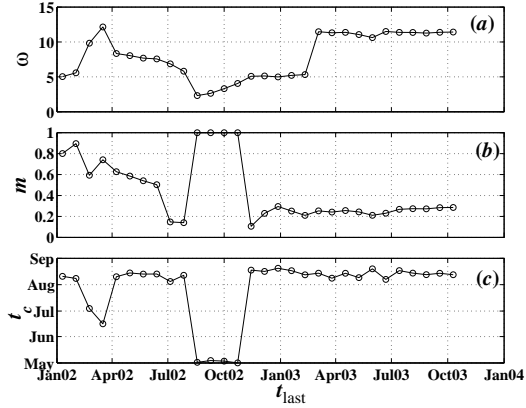}
\end{center}
\caption{Fitted parameters as functions of the end $t_{\rm{last}}$
of the fitted interval since the inception of the Chinese antibubble: (a)
$\omega$, (b) $m$, and (c) $t_c$.} \label{Fig:detectable}
\end{figure}

In this goal, we fit the SHSE composite index from 2001/08/09
to different ending date $t_{\rm{last}}$ with the first-order LPPL
formula (\ref{Eq:fit1}). Thirty ending dates
$t_{\rm{last}}$ have been used, taken equidistant in the time interval
from 2002/01/10 to 2003/10/28. This gives 30 fits and thus 30 groups of
parameters. Fig.~\ref{Fig:detectable} shows the three most
important parameters
$\omega$, $m$ and $t_c$ as functions of $t_{\rm{last}}$.

Panel (a) of Fig.~\ref{Fig:detectable} shows that the angular
log-frequency are stable after $t_{\rm{last}}={\rm{2003/03/04}}$
with $\omega=11.29\pm 0.25$, as ten $\omega$'s are in the interval
$(11,12)$ and one in the interval $(10,11)$. In the time interval
from 2002/11/14 to 2003/03/04, we find $\omega=5.15\pm 0.13$,
which is half the value for later $t_{\rm{last}}$. It thus seems
apparent that the fit has oscillated between locking in on the
fundamental or on its harmonic. The values of $m$ and $t_c$ are
also remarkably stable after 2002/11/14. We find $m=0.24\pm 0.04$
and $t_c={\rm{2001/08/14}}\pm 4$ (calendar days) for
$t_{\rm{last}}$ after 2002/11/14. If we use the following three
criteria for the qualification of the antibubble: (1) its angular
log-frequency $\omega$ should close to $\omega_f$ or its
harmonics, (2) its exponent $m$ should be less than $0.5$ (to
describe a significant power law decay), and (3) its estimated
critical time $t_c$ should be in August of 2001, then we conclude
that the antibubble could have been detected from the end of 2002.
We note that these three criteria are also met for
$t_{\rm{last}}=$ 2002/07/27 and $t_{\rm{last}}=$ 2001/08/12, two
dates which could have provided early precursors. These criteria
are similar to those used in advance predictions of crashes (see
\cite{SJ01Fei,S03}).

Thus, should we have detected the Chinese antibubble as early as 2002/07/27
before we started our own work in September 2002
on the worldwide antibubble that started in August 2000?
The answer is probably negative because such early detection would
not have been confirmed by the stability of the fitted parameters.
This example stresses the importance of verifying the stability
of parameters over successive time periods, before concluding
about the detection of a LPPL bubble or antibubble.
The 2002/07/27 alarm
occurred as an isolated case and could have been by chance.
We can however state with confidence that the existence of the antibubble
could have been confirmed by the end of 2002, based on the series
of confirmation of the three criteria from November 2002.
After 2003/03/04 when the fitted
$\omega$ is found close to $2\omega_f$, the confirmation of the
existence of the antibubble is even stronger because the occurrence
of precisely defined harmonics is a strong telling presence of a non-random signal.

\subsection{Predictability of market direction and optimal horizon}
\label{s2:OIH}

\begin{figure}[h]
\begin{center}
\includegraphics[width=7cm]{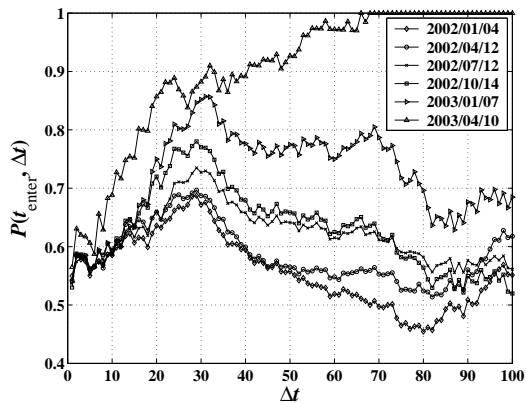}
\end{center}
\caption{Percentage $P(t_{\rm{enter}},\Delta{t})$
of successful predictions of the market direction for six different
beginning times $t_{\rm{enter}}$ for the application of our strategy
(explained in the text) and for various time horizons.} \label{Fig:OIH}
\end{figure}

An even better test for qualifying the Chinese LPPL antibubble is
to check if anomalous predictions can be obtained. We assess the
predictive power of the LPPL framework in the following way. We
define a fixed horizon $\Delta{t}$ and perform LPPL fits with
expression (\ref{Eq:fit1}) to the SHSE composite index (using
$I(t)=\ln[p(t)]$) from 2001/08/09 to a given time $t$. We then
extrapolate the fit from $t$ to $t+\Delta{t}$ using the fit and
thus obtain a prediction for the sign of the expected return
$r(t,\Delta{t}) = I(t+\Delta{t}) - I(t)$ over the next time period
$\Delta{t}$. We then perform the same operations for the SHSE
composite index up to the time $t+\Delta{t}$, and so on in time
steps of $\Delta{t}$ until the end date of 2003/10/28 is reached.
We are thus able to construct the percentage of times when the fit
predicts the correct market direction. We do this for different
starting times of the procedure, called $t_{\rm{enter}}$
(corresponding to a fictitious situation of a trader starting to
apply our strategy from $t_{\rm{enter}}$ till 2003/10/28).  We
have used six different $t_{\rm{enter}}$: 2002/01/04, 2002/04/12,
2002/07/12, 2002/10/14, 2003/01/07, and 2003/04/10. We also
perform the analysis of different time horizon $\Delta{t}$ from 1
day to 100 days. This gives us a bivariate percentage function
$P(t_{\rm{enter}},\Delta{t})$ of success, which is shown in
Fig.~\ref{Fig:OIH}.

For fixed time horizon $\Delta{t}$, the success percentage
$P(t_{\rm{enter}},\Delta{t})$ increases with $t_{\rm{enter}}$,
reflecting the increasing reliability of the fit with the
maturation of the antibubble. For fixed entering date
$t_{\rm{enter}}$, $P(t_{\rm{enter}},\Delta{t})$ has a clear
maximum at about 30 trading days (about six calendar weeks),
except for the most recent one for which
$P(t_{\rm{enter}},\Delta{t})$ increases monotonically with
$\Delta{t}$. This optimal $\Delta{t}$ for the first five
$t_{\rm{enter}}$ results from the competition between noise
(high-frequency price changes), which favors a larger $\Delta{t}$,
and the need to finely tune the fit to recent data, which favors a
smaller $\Delta{t}$. We note that the high predictability found
using our methodology does not translate in gains when negative
returns are predicted because shorting is forbidden in the Chinese
stock markets. The best that can be done is to exit the market and
keep one's wealth in a risk-free investment.

These findings have similarities and differences with those
obtained for the US market for a similar time period (see Fig. 7 of \cite{ZS03RG}).
In the case of the US S\&P 500 index,
$P(t_{\rm{enter}},\Delta{t})$ is an approximately increasing
function of $\Delta{t}$ and remains unchanged for different
$t_{\rm{enter}}$. However, it is interesting that $P(t_{\rm{enter}},\Delta{t})$
for the US markets also exhibits a locally maximum for
$\Delta{t}\approx 30$ days as for the Chinese stock market, suggesting
that the quality of the LPPL signal for prediction compared with the amount of
noise are similar. We also find a larger predictability (larger
$P(t_{\rm{enter}},\Delta{t})$) for not too large
$\Delta{t}$ in the Chinese compared with the US stock markets.

\subsection{What scenario for the future of the Chinese stock markets?}
\label{s2:FuturePred}

The cumulative evidence provided by our tests presented above
suggests to use our detection of an antibubble regime to predict
the future evolution of the Chinese stock market, similarly to the
advanced prediction in January 1999 for the Japanese Nikkei index
\cite{JS99IJMPC,JS00IJMPC} and the advance prediction in December
2002 for the US S\&P500 index \cite{SZ02QF}.

\begin{figure}[htb]
\begin{center}
\includegraphics[width=7cm]{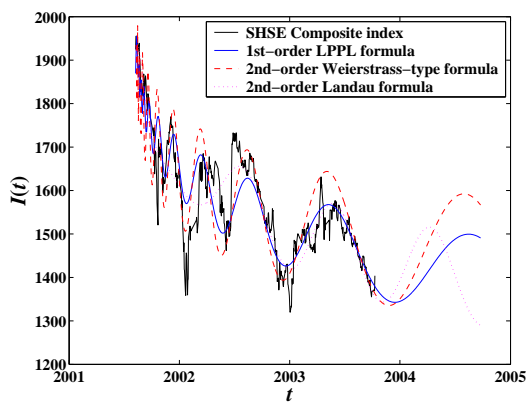}
\end{center}
\caption{(Color online) Forward prediction of
the Chinese stock market obtained by extrapolating the fits of the SHSE
composite data until 2003/10/28 using three different formulae
explained in the legend.} \label{Fig:FuturePred}
\end{figure}

In order to get the most robust and reliable advanced
prediction, we use the three different implementation of our LPPL theory
described above (first-order LPPL formula (\ref{Eq:fit1}),
LPPL with the second harmonic (\ref{Eq:fit2})
and second-order Landau expansion formula (\ref{Eq:Landau2})).
The advanced predictions are obtained by simply extrapolating the
fits with these three formulas performed using the data up to
2003/10/28. The results shown in
Fig.~\ref{Fig:FuturePred} predict that the
negative trend will bottom at the end of the year 2003. The
first-order LPPL formula and the second-order Weierstrass-type
formula predict that the rebound will end six or seven months
later, while  the second-order Landau formula is more pessimistic
in the sense that it predicts an earlier termination of the coming
bull market. In addition, the Landau formula (\ref{Eq:Landau2}) forecast a slightly
earlier recovery time.

However, there are two subtle issues we have to keep in mind.
First of all, our attempt to predict by extrapolating to the
future does not mean that we think the bearish market will
continue forever. Due to the large value of $\Delta_t$ (see
Sec.~\ref{s2:LPPLfit2Landau}) and the rather stable parameters
values (see Sec.~\ref{s2:Detactable}), the current antibubble has
not entered a phase shifting process and it is too early to detect
such phase shift \cite{ZS0310}. In addition, since the Chinese
stock markets are rather immature and heavily affected by
government policies \cite{C03}, the current antibubble may
terminate at some time in the future suddenly due to some
unexpected exogenous shock.

Let us add a cautionary note with respect to the predicted amplitudes.
As we can see in Fig.~\ref{Fig:FuturePred}, the dips are not well
captured by the fits which are by construction ``low frequency.''
Specifically, the two dips in January of
2002 and of 2003 were much deeper and more severe than described by the
fits. A similar dip could occur also close to the predicted
minimum corresponding to a plunge of the SHSE composite index
below 1300 sometimes in the remaining months of the year 2003.

\section{Universality versus idiosyncracy of the China antibubble}
\label{s1:UnivIdio}

\subsection{Comparison with other antibubbles} \label{s2:cmp}

The recognition of the present regime of the Chinese stock market
as an antibubble is tantamount to an attempt to classify universal
regimes in the dynamics of stock markets. It is thus important to
compare the 2001 Chinese stock market antibubble with those in the
other stock markets.

Several periods after 1990 have been classified as stock market
antibubbles:
\begin{enumerate}
\item the Japanese
Nikkei 225 antibubble started in January 1990 which lasted over
ten years \cite{JS99IJMPC};
\item the Argentina antibubble started
in June 1992 which lasted about six months \cite{JS00IJTAF};
\item the January 1994 antibubbles in Australia, France, Hong Kong,
Italy, New Zealand, Spain, Switzerland, and the United Kingdom
which lasted half to one year \cite{JS00IJTAF};
\item the Chile
antibubble which started in July 1995 and lasted about nine months
\cite{JS00IJTAF};
\item the September 1997 antibubbles in Russia
\cite{JSL99JR} and Venezuela \cite{JS00IJTAF} which lasted about
ten months;
\item the worldwide August 2000 antibubble in 23
indexes worldwide which has apparently not ended yet
\cite{SZ02QF,ZS02Global,GP03Polish}.
\end{enumerate}

Among all these cases, two in 1994 and 2000 involve several or many
countries simultaneously, suggesting the existence of
strong coupling and contagion mechanisms. All the others appear
to be localized to single markets. In the global 1994 antibubble,
the coupling was apparently delocalized across the different
markets with not clear single directing factor.
In the case of the 2000 antibubble, its seems that the
US market was and remains the leading factor affecting the other markets, as
witnessed by the extraordinary synchronization of other markets
with the US market (see figure 31 in \cite{ZS02Global}).

The current 2001 Chinese stock market antibubble occurs at a time
of a worldwide antibubble but remains by and large decoupled from
it. It seems that the Chinese antibubble is an example in which
idiosyncratic mechanism dominate over global coupling and
contagion. Let us compare further the 2001 China antibubble and
the 2000 global antibubble. We have previously investigated the
inter-dependence between the SHSE Composite Index and the US
S\&P500 Index including data up to December 2002, using several
measures of inter-dependence \cite{ZS02Global}: the
cross-correlation of weekly returns, linear regressions of indices
and of their returns \cite{R88FAJ}, a synchronization ratio of
joint occurrences of ups and downs \cite{ZS02Global} and an event
synchronization method recently introduced \cite{QKG02PRE}. The
China stock market was found to be uncorrelated or slightly
anti-correlated with the US market with a cross-correlation
coefficient of synchronous weekly returns of about -0.06 in a
three-month moving window. This should be compared with the west
European markets (France, Germany, Netherland, UK, {\it{et al.}})
which have a cross-correlation coefficient of about 0.3 with the
US market \cite{ZS02Global}. Comparing the Chinese antibubble
since August 2001 directly with the western antibubbles since
August 2000, we find that the daily return cross-correlation is
very low (-0.020 with the US S\&P 500 index, 0.008 with the
leading German index and -0.037 with the leading French index, to
list a few), while the return cross-correlation coefficients
between the western antibubbles are significantly higher (0.665
between US and Germany, 0.539 between US and France and 0.847
between Germany and France). In a nutshell, the Chinese stock
market has almost no correlation with the major western stock
markets despite their increasing mutual international trading.

From the view point offered by the LPPL formulas, the weak or
non-existent correlation between the Chinese antibubble
and the antibubbles of western markets can be attributed
to the conjunction of (1) different inception times $t_c$, (2) different
angular log-frequencies (and especially the existence of two angular
log-frequencies with similar amplitudes in the Chinese case) and
(3) different exponents
($\frac{1}{2}<m<\frac{3}{2}$ for most of the western markets compared
with $0<m<\frac{1}{2}$ for China).

\subsection{Particularities of the Chinese stock market}
\label{s2:WhyHow}

The fundamental reason underlying the special behavior
of the Chinese stock market compared with western markets
lies in its immaturity and the strong influence of the
government which still keeps a strong control.

In developed countries, profits and dividends are the
cornerstones of a fundamentally strong and mature stock market
\cite{F92}. The performance of a stock is strongly correlated with
the output of the company. Putting money into these market is
an investment more than a speculative gambling (except
during the late stage of speculative bubbles
\cite{white,Kindleberger,Garber,Shillerexu,S03}).
In contrast, in the
Chinese stock markets, activities are far from regularized. Many
companies attempt to collect money by issuing equities, while the
interest of the investors is not under careful consideration. Some
of the companies report inflated data of output in order to issue new
stocks and snatch more money (this is not
completely different from recent reported mispractice in the US
and elsewhere). This practice is particularly evident in the
extremely high initial public offerings (not unlike
the extreme IPOs of the late stage of the 2000 information
and technology bubble in the US). Therefore, the Chinese
stock market is more of a speculation market than an investment
market \cite{C03}. One could observe that the stock prices
soared during the bubble from May 1999 to June 2001
while the confidence of investors
decreases. The 1999-2001 bubble was thus even more unsustainable than
standard bubbles, and its inherent instability
eventually leads to a strong correction \cite{S03,S03PR}.

Institutional investors provide a significant fraction
of the capital invested the Western financial markets.
In the US as of November 2003, more than half of US households
are invested in mutual funds which hold \$7 trillion in assets,
almost half the total US market capitalization. Notwithstanding
recent scandals, mutual funds are supposed
to be more rational and to invest on longer term than
speculative agents, with positive stabilizing effects.
In contrast, a widely-accepted view is that the Chinese
stock markets are dominated by individual investors who are driven
by short term benefit making the markets highly volatile
\cite{C03,W01SSRN}. However, there are some dissonant views
arguing that the proportion of individual investors is exaggerated
and institutional investors including formal funds (a
proportion of 5\% of the tradable market capitalization by April
2001) and privately raised funds (a proportion of 43\% of the
tradable market capitalization by April 2001) occupy a large part
in the Chinese stock markets \cite{X01CJ,G03}. It is also argued
that privately raised funds have tighter ties with individual
investors \cite{X01CJ}. In general, the Chinese stock markets
respond to exogenous information, such as firm's accounting
information, stock exchanges announcement, government policies,
and so on \cite{CCS01JIAAT,GT01SSRN,CFG02JIFMA}.

\begin{figure}[h]
\begin{center}
\includegraphics[width=8cm]{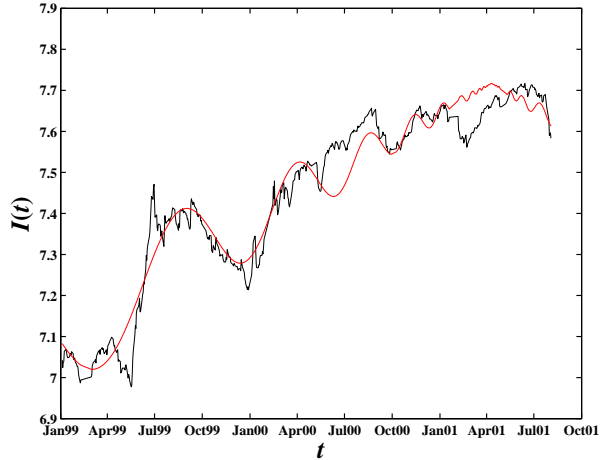}
\end{center}
\caption{Attempt to fit the (logarithm of the) SHSE Composite
Index from January 1999 to July 2001, using expression
(\ref{Eq:fit1}) with $\tau=|t_c-t|$, in order to test whether we
can detect a speculative bubble preceding the crash in July 2001.}
\label{bubbleattemp}
\end{figure}

During the development of the ``May 1999 - June 2001'' bubble,
several uncertainties coexisted that made investors rather
sensitive to government policies. The largest uncertainty was
related to the untradable shares. It has long been realized by
investors and by the Chinese government that the huge proportion
of the untradable shares blocks the heathy development of the
Chinese stock markets \cite{C03}. The problem is how and when to
reduce the holding of state-owned shares and eventually implement
full tradability. Heated discussions and debates on the ways to
address the problem marked the ``May 1999 - June 2001'' bubble
period until the crash in June 2001 following the historical high
on 2001/06/13. Figure \ref{bubbleattemp} shows an attempt to fit
the (logarithm of the) SHSE Composite Index from January 1999 to
July 2001, with expression (\ref{Eq:fit1}) with $\tau=|t_c-t|$, in
order to test whether we can detect a speculative bubble preceding
the crash in July 2001. As can be seen, the fit does not qualify a
clear LPPL signal because the exponent $m=1$ is large and the
oscillations are much more irregular than described by
log-periodicity (\ref{Eq:fit1}). According to the classification
of Johansen and Sornette \cite{classJS}, the crash that followed
this period should thus be characterized as ``exogenous.'' In
fact, the crash was indeed directly triggered by the promulgation
of the {\em{Tentative Administrative Measures for Raising Social
Security Funds through the Sale of State-Owned Shares}} by the
State Council of China\footnote{According to the {\em{Tentative
Administrative Measures for Raising Social Security Funds through
the Sale of State-Owned Shares}}, a joint stock company in which
the State holds an interest shall arrange for 10\% of the existing
State-owned shares to be sold as part of the offering, when
conducting an initial or secondary offering to the public. The
proceeds from the offering for sale will be contributed to
national social security funds.}. On 2002/06/24, the State Council
of China announced the removal with immediate effect of the
provisions of the {\em{Tentative Administrative Measures for
Raising Social Security Funds through the Sale of State-Owned
Shares}}. The evolution of the SHSE Composite Index from May 1999
- June 2001 is probably due to a combination of speculative
herding behavior and of policy-induced reactions, which is the
probable cause of the absence of a clear-cut LPPL signal. This is
one of those difficult cases where several processes intermix and
blur the endogenous versus exogenous nature of the process. The
state-owned share is still a dilemma for the Chinese policy-makers
and the Chinese stock markets and is expected to continue to
complicate the readings of the Chinese stock markets.

Steps in opening the stock markets to foreign investors are also
very recent. The first foreign institutional investor bought in
Chinese stock market no sooner than during the summer of 2003
(source: China Daily): UBS, one of the world's leading financial
service firms, became the first foreign financial institution to
invest in the US\$500 million A-share market in China. It placed
its first order under the QFII (qualified financial institutional
investor), which was previously open only to domestic investors.
The long-anticipated deal, billed by many as a major boost to the
market, failed to prop up China's flat yuan-currency A-share
market. This step is just the beginning of an interesting
development.

Our study shows that, notwithstanding all the idiosyncracies of
the Chinese markets, universal LPPL signatures reflect the
fundamental tendency of investors to speculate and herd.

\subsection{Direction of capital flow: impact of a growing real
estate bubble}

The total amounts of Chinese household saving deposits were
64,332.4 billion yuan by the end of 2000, 73,762.4 billion yuan by
2001 and 86,910.65 billion yuan by 2002 and reached 108,000
billion yuan by the end of September 2003 {\footnote{Monthly
historical data back to 1999 are available on the web site of the
People's Bank of China at
http://www.pbc.gov.cn/english/baogaoyutongjishuju/}. Over the long
run, higher savings tend to increase domestic investment
\cite{SD98} but is also channelled in the stock market and in real
estate. Indeed, there is not many alternative investments for
individuals, the two main investment channels being the stock
market and the real estate market.

Before the start of the 2001 antibubble, the increasing savings
and the wealth effect \cite{Wealtheffect} from market
capitalization gains fed investments in real estate. After the
crash and due to the continuing overall bearish nature of the
Chinese stock market, there is less new capital attracted to the
stock market. Available capital is thus finding other channels,
the main one being in real estate. In addition, China's property
market is boosted and is expected to continuously increasing in
the coming decade after the successes of Beijing's bid for the
2008 Olympic Games, of Shanghai's bid for the 2010 World Expo, and
of China's entry to WTO. Although there are controversies on the
existence of a real-estate bubble in China, it is accepted widely
that regional bubbles exist. In Zhuji City (Zhejiang Province),
which has a population of about 0.16 million people, the price of
a 120 M$^2$ house almost doubled over one year from September 2002
to September 2003 from \yen 165,000 RMB to about \yen 300,000 RMB.
In Xinzhuang of Shanghai, the house price in a same uptown roared
from \yen 2,800 /M$^2$ in November 2002 to \yen 5,400/M$^2$ in
October 2003. The Shanghai House Price Composite Index in October
2001 had a yearly growth of 27.8\%}. This growth rate is even a
little bit higher than in London, where an unsustainable
real-estate bubble may be in store \cite{ZS03PhysA}.

\begin{figure}
\begin{center}
\includegraphics[width=7cm]{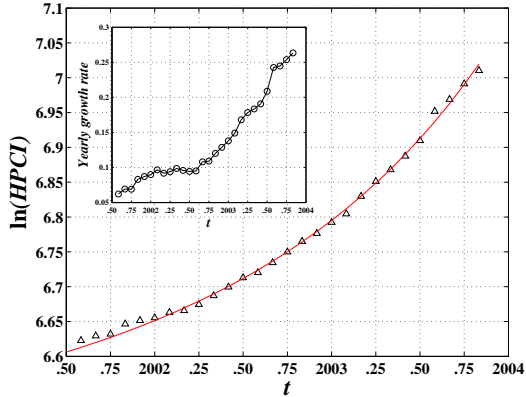}
\end{center}
\caption{Shanghai House Price Composite Index as a function of
time and its fit to (\ref{Eq:PL}) with $I(t)$ being the logarithm
of house price composite index. The parameters are respectively
$t_c=2008.80$, $m=-3.00$, $A=6.42$, and $B=74.1$. The inset gives
the yearly growth rate.} \label{Fig:house}
\end{figure}

Figure~\ref{Fig:house} shows the evolution of the Shanghai House
Price Composite Index for different months\footnote{Monthly
historical data back to August 2000 are available on the web site
of Shanghai Real Estate Appraisers Association at
http://www.valuer.org.cn/scjg/.} with its fit to a pure power law
(\ref{Eq:PL}). The inset shows the monthly growth rate. The fast
increasing growth of the growth rate means that, not only the
housing prices are growing fast, they are growing
super-exponentially: not only is the growth rate large, it is
growing and its growth is accelerating! These are the ingredients
for the future development a finite-time singularity
\cite{JSfinite,S03,STZ03}, announcing a tipping point and a strong
change of regime, possible a crash. However, since the estimated
critical time $t_c=2008.80$ is far from the present time, this
estimate is unreliable. It just implies that the house price will
continue to roar at a high speed and probably increasing
acceleration in the near future. We need to wait for more data to
accumulate before a better timing of the end of this speculative
bubble can be obtained. Further evidence of the existence of a
speculative real-estate bubble is found in the load and mortgage
reported by the People's Bank of China \cite{PBC03}. The
super-exponential acceleration of real-estate prices reflected in
the fast growth rate shown in Fig.~\ref{Fig:house} strongly
suggests that a real-estate bubble is on its way in Shanghai and
many other regions in China.

We should add that, in the recent two years, the Chinese
real-estate market has been fuelled by foreign capitals as well.
Foreign speculators's expectation of a {\em{Renminbi}}
appreciation led an inflow of dozens of billions of US dollars
(the ``hot'' money) in 2003. Part of the foreign capital are
deposits in China' banks and the remaining is invested in China.
If {\em{Renminbi}} appreciates, one can expect that the Chinese
stock and property markets will further fuel a bull phase, as
occurred in Japan in the second half of the 1980's. As the price
skyrockets, the positive feedback provided by the increasing
attraction of the real-estate market for foreigners will
eventually result in an unsustainable bubble
\cite{S03,S03PR,SZ03inflow}. Our previous work has shown in
several other contexts the strong positive feedbacks resulting
from foreign capital fluxes (see \cite{SZ03inflow} and references
therein).

\section{Concluding remarks}
\label{s1:concl}

In summary, we have identified an antibubble in the Chinese stock
market which started in August 2001 after the crash in June 2001.
The crash was probably the outcome of an instability
maturing during the development of a bubble from May 1999 to June
2001 and was triggered by a significant policy announcement.
It gave birth to the ensuing antibubble. This antibubble
exhibits the hallmarks previously identified in other antibubble
regimes in the last two decades, both in western and emergent
markets. These signatures comprise a power law relaxation
decorated with significant log-periodic undulations. The Chinese
antibubble exhibits universal as well as idiosyncratic features
when compared with other known stock market antibubbles. The
antibubble in China's equity market may be linked to the rapid
development of an ongoing property bubble in China, and
vice-versa. The real-estate bubble that we have clearly identified
is in turn fuelled by available capital from China's individual
investors and commercial banks as well as increasing influx for
foreign speculators. Our work shows that there is evident herding
behaviors among stock traders in China in spite of the fact that
the Chinese equity market is still immature and very much
influence by government policy. Actually, we have found maybe even
stronger imprints of herding that in other mature markets,
probably due to the immaturity of the Chinese market which seems
to attract short-term investors more interested in fast gains than
in long-term investments.

\bigskip
{\bf Acknowledgments:} We are grateful to Wen-Bin Mei who has
kindly provided the historical data of all the 14 indexes and to
Wei Deng for providing the Shanghai House Price Composite Index
data. This work was supported by the James S. Mc Donnell
Foundation 21st century scientist award/studying complex system.

\end{document}